\documentclass[fleqn,usenatbib]{mnras}

\usepackage{newtxtext,newtxmath}
\usepackage[T1]{fontenc}

\DeclareRobustCommand{\VAN}[3]{#2}
\let\VANthebibliography\thebibliography
\def\thebibliography{\DeclareRobustCommand{\VAN}[3]{##3}\VANthebibliography}
\usepackage{graphicx}	
\usepackage{amsmath}	
\usepackage{subfigure}

\title[MATLAS-42: Not Failed]{MATLAS-42, A Globular Cluster-Rich Ultra-Diffuse Galaxy That Diverges from the ``Failed Galaxy'' Formation Pathway}

\author[J. S. Gannon et al.]{Jonah S. Gannon$^{1}$\thanks{E-mail: jonah.gannon@gmail.com},
Duncan A. Forbes$^{1}$,
Francine R. Marleau$^{2}$,
Anna Ferr\'e-Mateu$^{3,4,1}$,
\newauthor
Aaron J. Romanowsky$^{5,6}$,
Maria Luisa Buzzo$^{7,1}$, 
and
Jean P. Brodie $^{1,6}$
\\
$^{1}$ Centre for Astrophysics and Supercomputing, Swinburne University, John Street, Hawthorn VIC 3122, Australia
\\
$^{2}$ Institute fur¨ Astro- und Teilchenphysik, Universitat¨ Innsbruck, Technikerstraße 25/8, A-6020 Innsbruck, Austria
\\
$^{3}$ Instituto de Astrof\'isica de Canarias, Calle V\'ia L\'actea S/N, E-38205, La Laguna, Tenerife, Spain
\\ 
$^{4}$ Departamento de Astrofísica, Universidad de La Laguna, 38206, La Laguna (S.C. Tenerife), Spain
\\
$^{5}$ Department of Physics and Astronomy, San Jos\'e State University, One Washington Square, San Jose, CA 95192, USA
\\
$^{6}$ Department of Astronomy \& Astrophysics, University of California Santa Cruz, 1156 High Street, Santa Cruz, CA 95064, USA
\\
$^{7}$ Astronomy Department, Yale University, 219 Prospect St, New Haven, CT 06511, USA
\\
}

\date{Accepted XXX. Received YYY; in original form ZZZ}

\pubyear{\the\year{}}

\begin{document}
\label{firstpage}
\pagerange{\pageref{firstpage}--\pageref{lastpage}}
\maketitle

\begin{abstract}
To date, there has been significant interest in globular cluster (GC)-rich ultra-diffuse galaxies (UDGs) and the evidence that they have formed via an unexpected, ``failed galaxy'' formation pathway. The majority of the evidence for ``failed galaxy'' UDGs originates from spectroscopic observations targeting passive GC-rich UDGs, with a focus on those residing in galaxy clusters. In this work, we study the gas-rich, GC-rich group UDG MATLAS-42 and derive its stellar population properties using the Keck Cosmic Web Imager. We measure a redshift for the galaxy ($V_{\rm R, \star}=2433\pm8$~km s$^{-1}$), confirming the previous assumptions that it is both part of the NGC~502 group and has an associated HI-reservoir ($V_{\rm R,HI}=2423\pm 15$~km s$^{-1}$). We measure integrated stellar populations and find the galaxy to be both young (mass-weighted age $=3.2^{+2.6}_{-1.5}$Gyr) and of average-to-low metallicity ($[M/H]=-1.19^{+0.42}_{-0.30}$ dex). When considering these properties in the context of the galaxy's formation, we note it likely does not follow the ``failed galaxy'' formation pathway commonly attributed to GC-rich, cluster UDGs, as it has experienced recent star formation. At most it started failed, however, it has recently rejuvenated its star formation. Finally, we build a toy model of the passive evolution of this galaxy, finding that its relative GC-richness (i.e., $M_{\rm GC}/M_\star$) will likely decrease with time as GCs slowly evaporate/disrupt to contribute to the stellar mass of the galaxy. Due to this, we hypothesise that it is likely not a low redshift analogue of the progenitor to a ``failed galaxy'' UDGs.
\end{abstract}




\section{Introduction}
Dwarf galaxies are expected to dominate the number density of galaxies in the Universe at all epochs \citep{Wright2017, Martin2019} and, as galaxy stellar mass increases, the fraction of stellar mass comprised of accreted dwarf galaxies also increases (see e.g., \citealp{Remus2022}). The advent of deep, wide-area surveys is increasingly finding large numbers of previously overlooked dwarf galaxies. For example, \citet{vanDokkum2015} found 47 large, low surface brightness ``ultra-diffuse galaxies'' (UDGs) in the Coma cluster. These galaxies were characterised by having large half-light radii ($>1.5$~kpc) and low surface brightnesses ($\mu_{g,0}>24$~mag arcsec$^{-2}$), which placed them in a region of parameter space that was previously sparsely populated.

Since their discovery, UDGs have been found in all environments, i.e., from the field to massive galaxy clusters, and many large surveys have been conducted to study their formation (e.g., SMUDGes; \citealp{Zaritsky2019}; MATLAS \citealp{Marleau2021}; LEWIS \citealp{Iodice2023}; EUCLID-ERO Perseus \citealp{Marleau2025}). Their study has posed many puzzles that are yet to be solved. Namely, a consensus has yet to be reached as to how they form, with galaxy formation simulations suggesting that stellar feedback \citep{DiCintio2017, Chan2018}, higher than average halo spin \citep{Amorisco2018, Liao2019}, galaxy mergers \citep{Wright2017}, high velocity galaxy collisions \citep{Lee2021, Lee2024}, ram pressure stripped material \citep{Ivleva2024}, tidal heating and tidal stripping \citep{Carleton2019, Sales2020}, may be the dominant mechanisms that form UDGs. Combinations of these processes (e.g., \citealp{Jiang2019}) help to explain the full observed properties of the UDG population (see e.g., \citealp{FerreMateu2023}). 

Crucially, none of the proposed formation pathways that originate from galaxy formation simulations are able to explain how some UDGs are observed to be extremely globular cluster (GC)-rich for their stellar mass (e.g., \citealp{vanDokkum2017, Forbes2020, Danieli2022, Saifollahi2022, Saifollahi2025, Li2025}). In order to create these galaxies, a separate formation pathway, that of a ``failed galaxy'' or ``pure stellar halo'' \citep{vanDokkum2015, Peng2016, Danieli2022, Forbes2024, Forbes2025} has been proposed. In this scenario, the UDG forms rapidly at high redshift alongside its GC system. It then quenches and passively evolves until observed today at low redshift. As of yet, no cosmological simulation has found a UDG to follow this pathway; however, their possible properties have been studied using a toy model and the MAGNETICUM simulations \citep{Gannon2025a}. Despite their lack of simulated counterparts, there exists myriad evidence for this formation pathway for UDGs:

\begin{enumerate}
    \item Their large GC systems (and hence ratio of GC mass to stellar mass; $M_{\rm GC}/M_\star$) imply a halo more massive than the stellar mass halo mass relationship (\citealp{Forbes2024}; \citealp{Forbes2025b}; Dornan \& Harris in press.).

    \item Many UDG stellar velocity dispersions agree with their residing in massive dark matter halos \citep{Gannon2022, Zaritsky2023, Forbes2024}. This reinforces the evidence for massive halos from point (i). 

    \item Many UDGs are old, with quick single-burst star formation histories and alpha-enhancement indicative of an early, fast formation \citep{FerreMateu2018, FerreMateu2023}. A halo evolved through catastrophic quenching and passive stellar evolution from high redshift would help explain its high stellar to halo mass ratios \citep{Gannon2025a}.

    \item Many UDG stellar population gradients are flat with radius, indicative of an ``everything everywhere all at once'' formation pathway, which is expected for ``failed galaxies'' \citep{Villaume2022, FerreMateu2025}. This is likely related to their formation from single burst star formation events (point iii).

    \item Many UDGs in the spectral energy distribution (SED) studies of \citet{Buzzo2024, Buzzo2025} were found to have low metallicities for their stellar mass and to be better aligned with a high redshift mass -- metallicity relationship. This helps to confirm their high redshift formation.

    \item Finally, \citet{Buzzo2025} used the K-means unsupervised clustering algorithm to assess if multiple populations of UDGs reside in their sample data. They found two populations, with one resembling an extension of normal dwarf galaxies and the other resembling more the properties of ``failed galaxy'' UDGs. This finding suggests that the ``failed galaxy'' formation pathway for UDGs is distinct from other dwarf formation mechanisms.
\end{enumerate}

Crucially, however, much of the above evidence for the ``failed galaxy'' pathway of UDG formation comes from studies which have a known bias to UDGs residing in dense, cluster environments \citep{Gannon2024b}. To further test our understanding of this unique and interesting pathway of galaxy formation, we must increase the sample of GC-rich UDGs in less dense environments. 

The MATLAS Survey performed deep, wide-field optical imaging of nearby massive lenticular and elliptical galaxies. Included in their deep imaging are myriad dwarf galaxies, many of which fit the UDG criteria \citep{Habas2020, Marleau2021, Poulain2021}. Many of these dwarf galaxies have now received follow-up HI observations \citep{Poulain2022} and dedicated spectroscopy from MUSE \citep{Muller2020, Heesters2023}. Of interest to this work is the study of \citet{Marleau2024}, which used \textit{Hubble Space Telescope} (HST) imaging of 44 UDGs and 30 NUDGes (Nearly UDGs)\footnote{Originally this work had 38 UDGs and 36 near UDGs; however, their surface brightnesses have recently been revised in Marleau et al. (Erratum Submitted).} in low-density environments to study their GC systems. The vast majority of the UDGs they studied (64\%) contained few, if any, GCs. Indeed, only a small population, 9\%, were found to be particularly GC-rich. One of these UDGs, NGC5846\_UDG1, a.k.a. MATLAS-2019, has already been the target of extensive study due to its rich GC-system \citep{Forbes2019, Danieli2022, Muller2020, Muller2021, Forbes2021, FerreMateu2023, Marleau2024, Haacke2025}.

Also among the small sample of GC-rich UDGs observed by \citet{Marleau2024} was MATLAS-42. MATLAS-42 is near unique for GC-rich UDGs in that it also has a detection of HI gas, likely to be associated with the galaxy \citep{Poulain2022}. In this work, we perform Keck Cosmic Web Imager (KCWI) spectroscopy of MATLAS-42 to analyse its spectroscopic properties and compare them to the GC-rich UDGs that have been previously studied. A proposed pathway to quench ``failed galaxy'' UDGs is that feedback from the GC formation ejects the gas, leading to a natural question as to why MATLAS-42 has gas while other GC-rich UDGs do not. We will study how MATLAS-42 may have formed and if it also follows a ``failed galaxy'' formation pathway despite its location in a lower density environment and its retention of a significant gas reservoir. 

In Section \ref{sec:pao}, we summarise the properties of MATLAS-42 from the literature. We also present our new KCWI observations of the galaxy and the results of these new observations. In Section \ref{sec:discussion}, we discuss the properties of MATLAS-42 in the context of previously spectroscopically observed UDGs and its possible formation pathways. Further, we discuss how MATLAS-42 may evolve in time if left to passive evolution. A brief summary and our conclusions are presented in Section \ref{sec:conclusions}. Magnitudes are in the AB system unless otherwise stated. They have been extinction corrected with an extinction of 0.11 in $g$.

\begin{table}
    \centering
    \begin{tabular}{ll}
    \hline
    MATLAS-42 &  \\ \hline
    RA [J2000] & 20.6260\\
    Dec. [J2000] & +8.7614  \\
    $D$~[Mpc] & 33.46 \\
    Host & NGC 502  \\
    $N_{\rm GC}$ & $22\pm7$\\
    $\log(M_{\rm HI} [M_\odot])$  & $8.18\pm0.06$ \\
    $V_{\rm R, HI}$ [km s$^{-1}$] & $2423 \pm 15$\\ \hline
    $m_{g}$ [mag] & 17.46  \\
    $R_{\rm e}$ ['']/[kpc] & 19.44/3.11 \\
    $b/a$ & 0.6 \\
    PA [deg] & $-$43 \\
    S\'ersic $n$ & 0.45  \\
    $\langle\mu_{g}\rangle_{\rm e}$ [mag arcsec$^{-2}$] & 25.94\\
    $\log(M_{\star}/$M$_\odot)$ & 8.17$^{+0.08}_{-0.1}$ \\
    $M_\star/L_V$ & 0.97 \\ \hline
    $V_{\rm R, \star}$ [km s$^{-1}$] & 2433$\pm$8 \\
    $[M/\mathrm{H}]$ [dex] & $-1.19^{+0.42}_{-0.30}$ \\
    Age [Gyr] &  $3.2^{+2.6}_{-1.5}$ \\ \hline
    \end{tabular}
    \caption{The key properties of MATLAS-42 from the literature. The properties before the first divide are taken from the MATLAS Survey. In particular, we take data from their online data retrieval tool along with \citet{Poulain2021, Poulain2022} and \citet{Marleau2024}. The properties in the middle divide are from the imaging and SED fitting of \citet{Buzzo2024}. The properties in the final divide are from this work.}
    \label{tab:target_summary}

\end{table}

\section{MATLAS-42 Properties and Observations} \label{sec:pao}
\begin{figure}
    \centering
    \includegraphics[width=0.45 \textwidth]{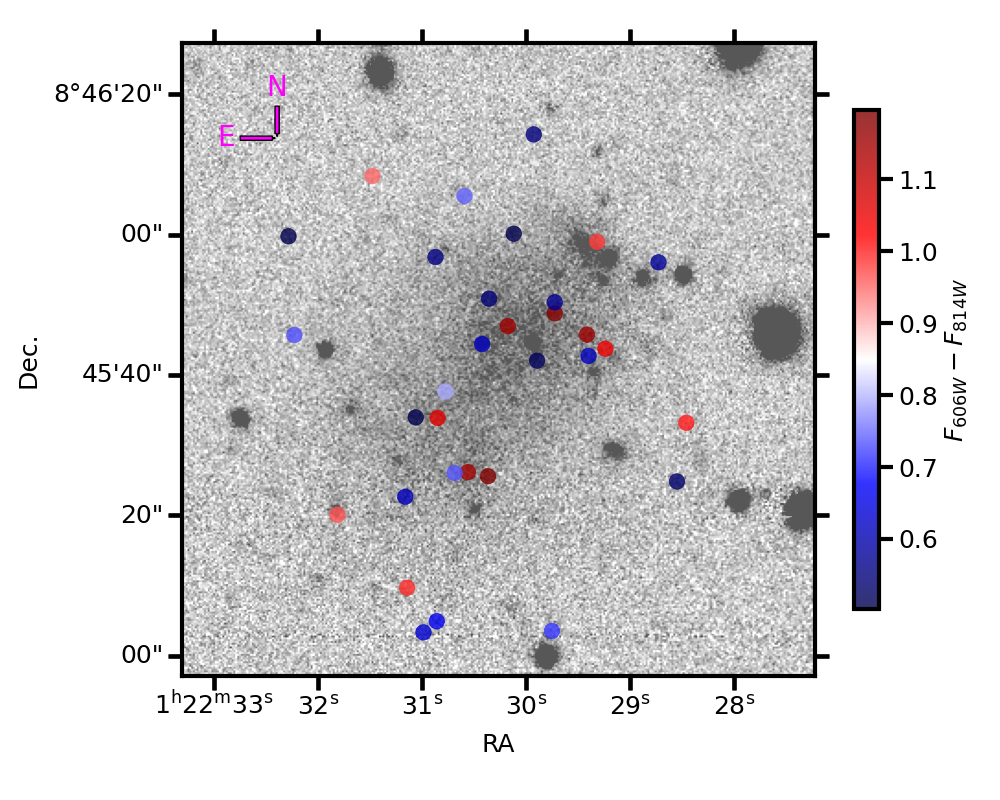}
    \caption{The projected distribution of globular cluster candidates around MATLAS-42 on the sky. The cutout is 1.5$\times$1.5' (14.6$\times$14.6~kpc)}. Candidates are colour coded by their $F_{606W} - F_{814W}$ colour from the \textit{HST} imaging of \citet{Marleau2024}. North and East are as indicated. The bluer and redder populations of GCs are approximately evenly distributed around the galaxy, with a small extension of both populations coinciding with the irregular extension of MATLAS-42's stellar body to the south-east.
    \label{fig:2d_colour}
\end{figure}

\begin{figure}
    \centering
    \includegraphics[width=0.5\textwidth]{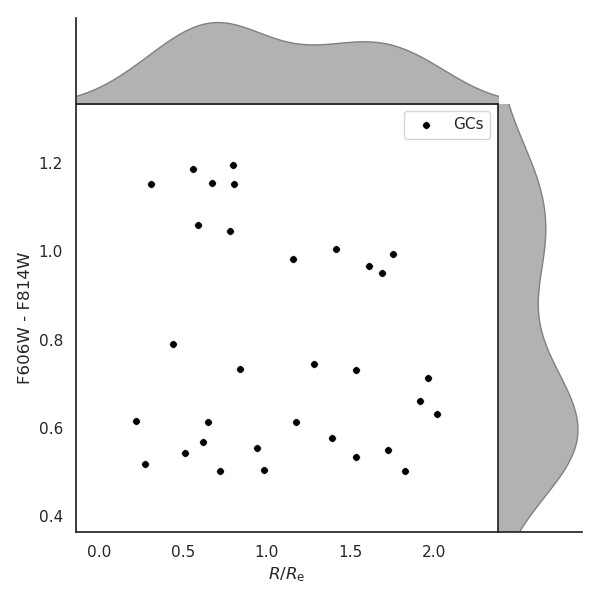}
    \caption{GC system candidate colour \textit{vs} projected radius from the galaxy centre. Kernel density estimates of each axis property are plotted alongside each axis. We take our GC positions and colours from the study of \citet{Marleau2024}, which used Vega magnitudes. Note that a significant fraction of the 32 plotted candidates, likely about a third, are statistical contaminants. While GC candidates do not show a strong trend with radius, there is a bimodality in colour. }
    \label{fig:GCproperties}
\end{figure}

\begin{figure*} \label{fig:figure_1}
    \centering
    \begin{subfigure}
        \centering
        \includegraphics[width = 0.3 \textwidth]{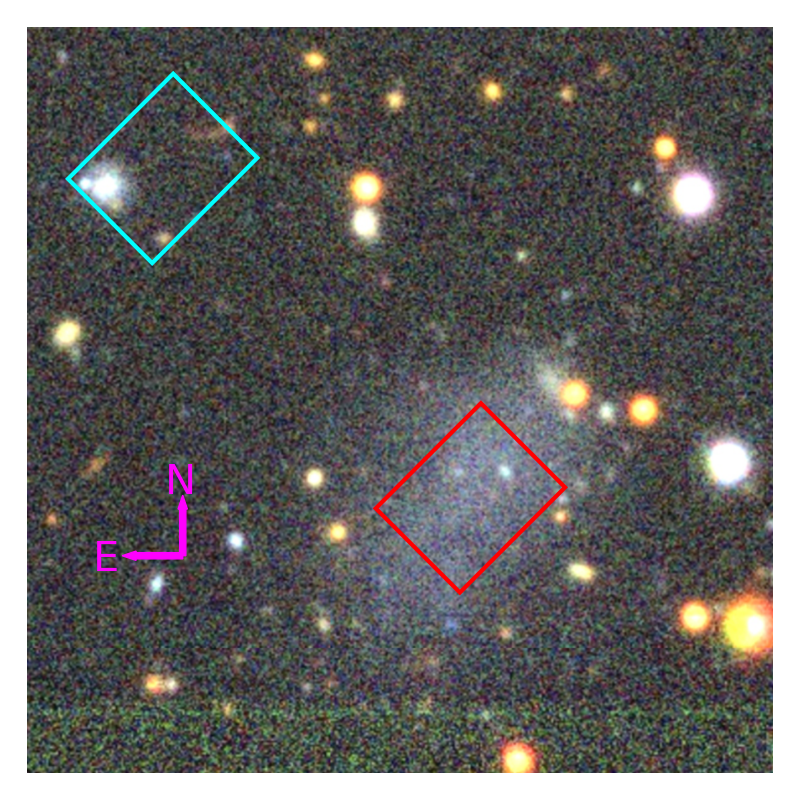}
    \end{subfigure}
    ~ 
    \begin{subfigure}
        \centering
        \includegraphics[width = 0.6 \textwidth]{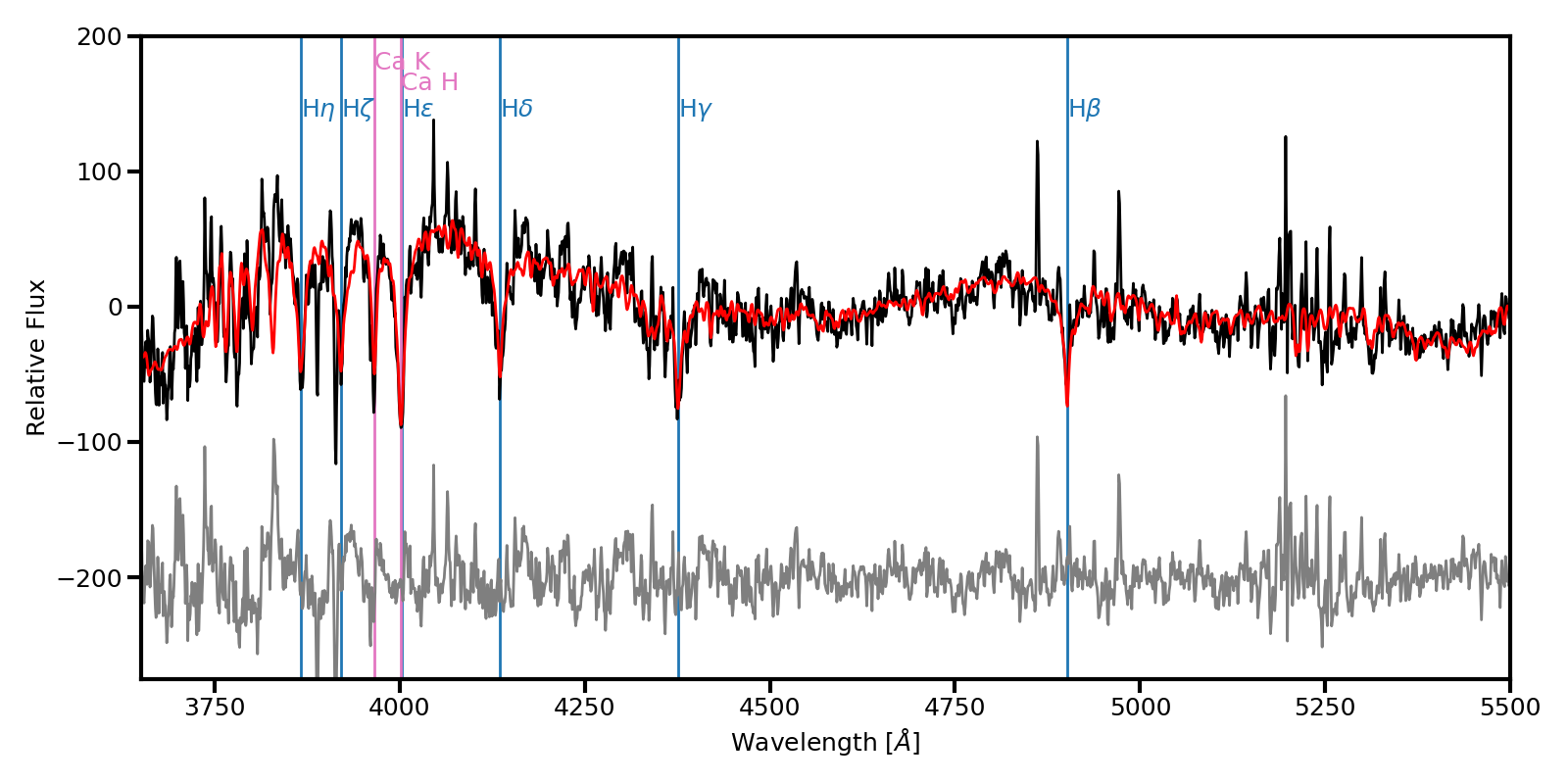}
    \end{subfigure}
    \caption{\textit{Left:} A 2'$\times$2' cutout centred around MATLAS-42 taken from the DECaLS Legacy Skyviewer (available \href{https://www.legacysurvey.org/viewer?ra=20.6258&dec=8.7614&layer=ls-dr9&zoom=16}{here}). In red, we show the positioning of KCWI observations. In cyan, we show the offset sky. Included in the offset sky is a small background galaxy. North is up and east is left as indicated. At MATLAS-42 distance, the 16''$\times$20'' medium slicer of KCWI corresponds to $\sim2.8\times3.5$~kpc. \textit{Right:} Our spectrum for MATLAS-42 (black) with best fitting \texttt{pPXF} template (red) and with residuals of the fit also shown (grey). A clear Balmer series is evident (blue vertical lines) making it likely that this is a young galaxy -- as confirmed by our \texttt{pPXF} fitting. }
\end{figure*}

In this Section we introduce the properties previously established for MATLAS-42 in the literature and the additions we make to these known properties. We describe the data we gathered and the methods we use to analyse it along with the results of these methods. 

\subsection{Literature} \label{subsec:literature}
We summarise the key literature properties of MATLAS-42 in Table \ref{tab:target_summary}. Much of these data are taken from the MATLAS survey \citep{Marleau2021}. In particular, we have used their online tool\footnote{\href{https://matlas.astro.unistra.fr/class/MATLAS_dwarfs_individual.php?id=MATLAS-42}{https://matlas.astro.unistra.fr/class/MATLAS\_dwarfs\_individual.php \allowbreak ?id=MATLAS-42}} to retrieve data, which largely comes from the works of \cite{Marleau2021, Marleau2024} and \citet{Poulain2021, Poulain2022}. We additionally take the spectral energy distribution (SED) fitting results for MATLAS-42 from \citet{Buzzo2024}, which we will compare to our spectroscopic fitting results later in this work. 

A few properties of MATLAS-42 are of particular note for this paper. Namely:

\begin{enumerate}
    \item MATLAS-42 has a half-light radius of 19.44'' corresponding to 3.11 kpc at the assumed distance of the NGC~502/NGC~524 group, 33.46 Mpc \citep{Buzzo2024}. Combined with its low surface brightness, $\langle\mu_{g}\rangle_{\rm e} = 25.94$~mag arcsec$^{-2}$ \citep{Buzzo2024}, this galaxy is a UDG \footnote{$\langle\mu_{g}\rangle_{\rm e} \approx \mu_{g, 0} +1.123$ for $n=1$ \citep{Graham2005}. Hence $\langle\mu_{g}\rangle_{\rm e} = 25.123$ mag arcsec$^{-2}$ is approximately the same as $\mu_{g, 0} =24$ mag arcsec$^{-2}$ and thus MATLAS-42 meets the original UDG criteria.} once the assumption of its distance is verified (as it will be in Section \ref{sec:ppxf}). 

    \item MATLAS-42 has a total GC system of 22 $\pm$ 7 GCs \citep[][see further for details of methodology on how this total GC estimate was derived]{Marleau2024}. This places it as ``GC-rich'' under the definition of \citet{Gannon2022}, which separates UDGs into GC-rich/GC-poor based on whether they host over/under 20 GCs.  The measured GC system of MATLAS-42 corresponds to having a $M_{\rm GC}/M_\star\approx$2.9\% assuming an average GC mass of $2\times10^5$~M$_\odot$. This is high fraction of stellar mass in its GC system, atypical for a UDG in a lower density group/field environment (see e.g., \citealp{Jones2023}), with the majority of GC-rich UDGs thus far being found in galaxy clusters. Having $\sim22$~GCs also implies a relatively massive halo mass of log$(M_{\rm Halo} / \mathrm{M_\odot})=10.87\pm14)$ \citep{Marleau2024}.

    \item MATLAS-42 has a coincident HI detection from the ALFALFA survey \citep{Haynes2018, Poulain2022}, which implies gas associated with the galaxy. The vast majority of UDGs studied spectroscopically thus far are in dense environments and without notable gas content. Furthermore, it is not expected for a dwarf galaxy with high $M_{\rm GC}/M_\star$~to have significant gas content at late times, as this gas content would likely lead to star formation after the formation of GCs, decreasing its richness relative to its stellar mass.

    \item It is also of note that MATLAS-42 is both irregular in morphology, with a second stellar component to its south-east (see Figure \ref{fig:2d_colour}), and has a close companion in MATLAS-41 located to the North West of the UDG \citep{Marleau2024}. The second stellar component is well traced by an extension of the GC candidate population in that direction (see Figure \ref{fig:2d_colour}/Figure B1 of \citealp{Marleau2024}). 
    
    \item The galaxy is slightly bluer than the other UDGs in MATLAS \citep{Marleau2024}, however, it is not extremely blue. There is a hint of GALEX emission associated with the galaxy. Both are consistent with relatively recent star formation. 

    \item Finally the stellar populations of MATLAS-42 have been studied previously using SED fitting \citep{Buzzo2024}. They measured a metallicity of $[M/\mathrm{H}]=-1.34^{+0.30}_{-0.19}$~dex and an age $= 5.97^{+5.46}_{-3.59}$~Gyr.

\end{enumerate}

With regards to the GC system, it is worth noting that, despite the $\sim1/3$ statistical contamination rate of the GC candidates \citep{Marleau2024}, there exists a clear colour bimodality, shown in Figure \ref{fig:GCproperties}. The bluest population of GC candidates is distributed evenly with radial distance out to two half-light radii (See also Figure \ref{fig:2d_colour}). 

The colour bimodality is also uncommon for a UDG as their GC populations are largely found to have unimodal (blue) colours (e.g., \citealp{Janssens2024}). \citet{Fielder2024} have shown a few examples of tidally affected UDGs that have colour spans in the range $0.5<V-I<1.5$, approximately similar to what is observed for MATLAS-42 here. We will consider tidal influences on MATLAS-42's formation later in Section \ref{sec:formation}.

Finally, it is worth stating that while MATLAS-42 has been associated with the lenticular galaxy NGC~502, NGC~502 itself has been shown to be a likely member of the NGC~524 group \citep{Brough2006}. This may provide evidence of the galaxy being closer than the assumption of association with NGC~502 would allow. Here, we keep the assumption of NGC~502's distance as it allows the best comparison with SED fitting results previously obtained by \citet{Buzzo2024}. The largest effects of a change of distance would occur for the galaxy's measured GC system, as the distance was assumed as part of the methods of \citet{Marleau2024} in estimating its rich GC system and in its estimated total stellar mass. It is beyond the scope of this work to repeat their analysis at various distances to evaluate this change. Overall, however, our spectroscopically measured properties (e.g., age, metallicity) are distance independent, and so a change of distance would not largely change our measurements of the galaxy's properties below. We thus provide the caveat to our work by stating we assume the same distance as others who have studied the galaxy thus far (33.46 Mpc).

\subsection{KCWI Data}
This work makes use of data from the integral field spectrograph KCWI on the Keck II telescope \citep{Morrissey}. The data were observed on the night of 2023, November 12th. Conditions were clear with 1.1'' seeing. The blue-arm of KCWI was configured using the Medium slicer and BL grating with a central wavelength of 4550\AA. In this configuration, it has an instrumental spectral resolution of 2.53\AA~(FWHM). Our observations were post-KCWI upgrade where the red arm was simultaneously available however, these data did not prove to have sufficient S/N for analysis and will not be discussed further. In total, there were 10$\times$1320 exposures targeting the galaxy, with 6$\times$1320 exposures targeting an offset sky position. The target positioning of both is shown in Figure \ref{fig:figure_1} (\textit{Left}). We note that our accuracy in returning to the galaxy position was frequently off by $\sim$3''. This was due to a power outage midway through observing, necessitating a change of telescope control, affecting our accuracy in returning to the galaxy position when swapping between sky and galaxy exposures. Despite the offset, the large size of MATLAS-42 still allowed these exposures to reside on the galaxy. 

These data were reduced with the standard Keck pipeline without sky subtraction before being cropped to a common wavelength range. These data were then stacked using the routine \texttt{MOSAIC}\footnote{\href{http://montage.ipac.caltech.edu/}{http://montage.ipac.caltech.edu/}}, and a spectrum was extracted, masking compact sources and weighting by the weight map of the stacked data. This spectrum was sky subtracted using the PCA sky subtraction routine described in \citet{Gannon2020} and the offset sky spectra. Our final, reduced spectrum is shown in Figure \ref{fig:figure_1} (\textit{Right}) and has a S/N of 12 \AA$^{-1}$. While there is a clear Balmer series, there are no apparent emission lines, indicating that it is not currently forming stars.

\subsection{\texttt{pPXF} Fitting} \label{sec:ppxf}
In order to extract stellar population information we fit our spectrum using the full spectral fitting routine \texttt{pPXF} \citep{Cappellari2004, Cappellari2017, Cappellari2022}. As templates, we use E-MILES models \citep{Vazdekis2015} based on a \citet{Kroupa} IMF and the BASTI isochrones \citep{Pietrinferni2021}. These templates have a spectral resolution of 2.52 \AA~(FWHM) which is well-matched to our data and cover a metallicity range of $-2.27\le[M/H]\le+0.4$~dex and an age range of 0.03 to 14 Gyr. There are 53 steps in age and 12 steps in metallicity, creating 636 distinct combinations of age and metallicity in these models. We fit our data using a 15th-order multiplicative polynomial and a 1st-order additive polynomial to help correct for any flux calibration errors that affect the shape of the galaxy's continuum. Our fitting includes gas templates but does not include \texttt{pPXF}'s regularisation parameter, which has been shown to have a negligible effect for UDGs \citep{FerreMateu2018, FerreMateu2023}. Our fitting included \texttt{pPXF}'s \texttt{clean} parameter to help remove any sky line residuals present in the spectrum (such as those evident near 5250~\AA). After an initial fit, we run 10000 bootstrap fits following the \texttt{pPXF} example \texttt{ppxf\_example\_population\_bootstrap.ipynb}\footnote{\href{https://github.com/micappe/ppxf_examples/blob/main/ppxf_example_population_bootstrap.ipynb}{https://github.com/micappe/ppxf\_examples/blob/main/\allowbreak ppxf\_example\_population\_bootstrap.ipynb}.}. Final stellar population results are quoted from the median of these fits with 16th and 84th percentiles as the uncertainties. Using this process we have measured a mass-weighted age of $3.2^{+2.6}_{-1.5}$ Gyr and a mass-weighted metallicity $[M/H]$ = $-1.19^{+0.42}_{-0.30}$ dex. Interestingly, while these results are stable and well-constrained, we were unable to derive a reliable star formation history for MATLAS-42. We discuss this further in Appendix \ref{app:fitting}.

From our initial \texttt{pPXF} fit, we also derive a recessional velocity, 2433$\pm$8 km s$^{-1}$, which has been corrected to the barycentre. We note that this redshift confirms the association of MATLAS-42 with the NGC~502/NGC~524 group \citep[$2524\pm5$ km s$^{-1}$;][]{Cappellari2011} as assumed by both the MATLAS survey and \citet{Buzzo2024}. A confirmation of the distance to MATLAS-42 confirms its status as a UDG. It also confirms its association with the HI gas detected in the ALFALFA survey, which has a recessional velocity of 2423$\pm$15 km s$^{-1}$ \citep{Poulain2022}, within our stellar recessional velocity's uncertainties. Our stellar population results show that MATLAS-42 is both young and moderately metal-poor, summarised in Table \ref{tab:target_summary}. Furthermore, our age and metallicity are in agreement with those from the SED fitting of \citet{Buzzo2024}. Their measured metallicity was $[M/\mathrm{H}]=-1.34^{+0.30}_{-0.19}$~dex and their age $= 5.97^{+5.46}_{-3.59}$~Gyr. 

Finally, we are also able to fit a recessional velocity for the galaxy included in the offset sky position. Here, we derive a recessional velocity of 41852$\pm$5 km s$^{-1}$. This galaxy is therefore in the background of the group and will not be considered further in this work. 

\section{Discussion} \label{sec:discussion}

\subsection{The Mass -- Metallicity and Age -- Metallicity Relations}

\begin{figure}
    \centering
    \includegraphics[width=0.48\textwidth]{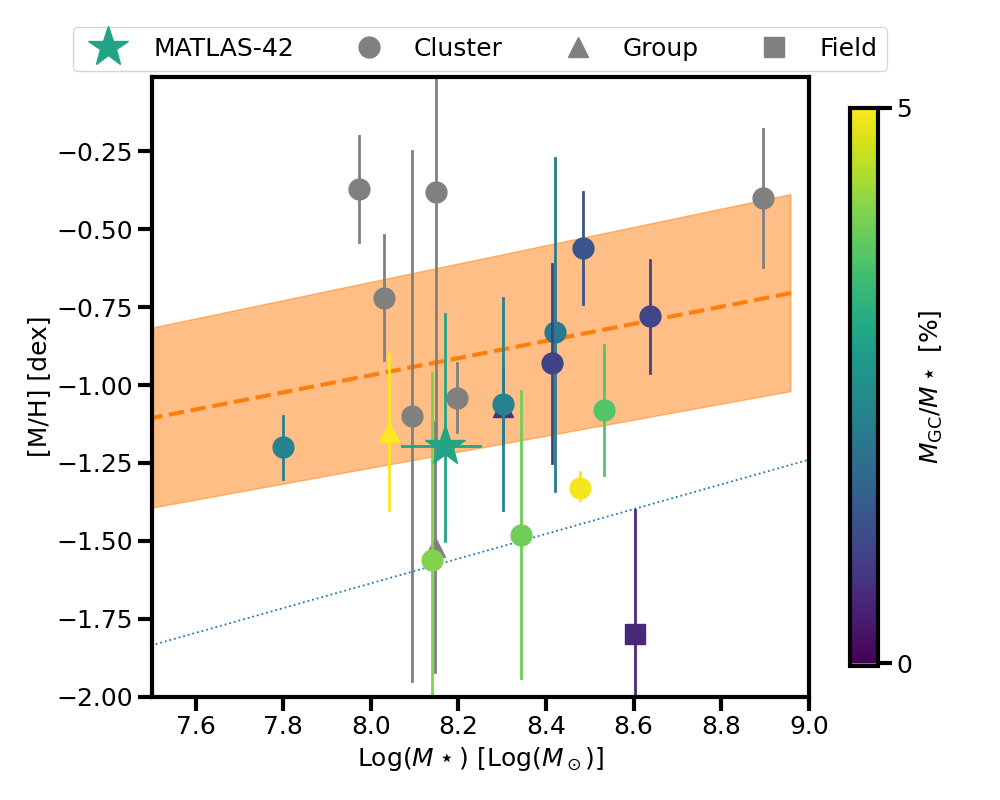}
    \caption{The stellar mass--metallicity relationship. We include MATLAS-42 (star) along with UDGs from the catalogue of \citet[][points]{Gannon2024b} where most come from the study of \citet{FerreMateu2023}. The stellar mass -- metallicity relationship for classical dwarfs at $z=0$~is plotted from \citet[][orange dashed line with 1-$\sigma$ shading]{Kirby2013} along with the simulated relationship at $z=2.2$~from \citet[][blue dotted line]{Ma2016}. Points are colour-coded by their relative GC richness with their style corresponding to their environment (circles are in clusters, triangles are in groups and squares are located in the field). UDGs where no GC count is available are plotted in grey.  MATLAS-42 is slightly more metal-poor than the stellar mass -- metallicity relationship; however, it is not as metal-poor as the expectation at high redshift. It is a crucial addition to the number of lower-density environment UDGs studied so far.}
    \label{fig:mass_met}
\end{figure}

In Figure \ref{fig:mass_met}, we show the location of MATLAS-42 in the mass -- metallicity relationship. The established relationship for classical dwarfs at $z=0$~is plotted from \citet{Kirby2013} along with the simulated relationship at $z=2.2$~from \citet{Ma2016}. We additionally show UDGs from the spectroscopic catalogue of \citet{Gannon2024b}\footnote{Retrieved from \href{https://github.com/gannonjs/Published_Data/tree/main/UDG_Spectroscopic_Data}{https://github.com/gannonjs/Published\_Data/tree/main/\allowbreak UDG\_Spectroscopic\_Data} on January 20th, 2025, see full references in the Data Availability section.}. The plot is colour-coded by the UDG's $M_{\rm GC}/M_\star$ and point styles are per their environment as listed in the \citet{Gannon2024b} catalogue. MATLAS-42 largely follows the $z=0$ stellar mass--metallicity relationship, being within the scatter to the metal-poor side. It is less compatible with the relationship at higher redshift. Many UDGs in the spectroscopic catalogue of \citet{Gannon2024b}, which (for stellar populations) primarily come from the work of \citet{FerreMateu2023}, are of similar mass, metallicity and GC richness. MATLAS-42 is unusual, however, as it is in a group environment. The majority of UDGs that have been studied thus far are in a denser, cluster environment. Its stellar populations (mass and metallicity) are most similar to the other UDGs that have been previously studied in a group environment.

\begin{figure}
    \centering
    \includegraphics[width=0.48\textwidth]{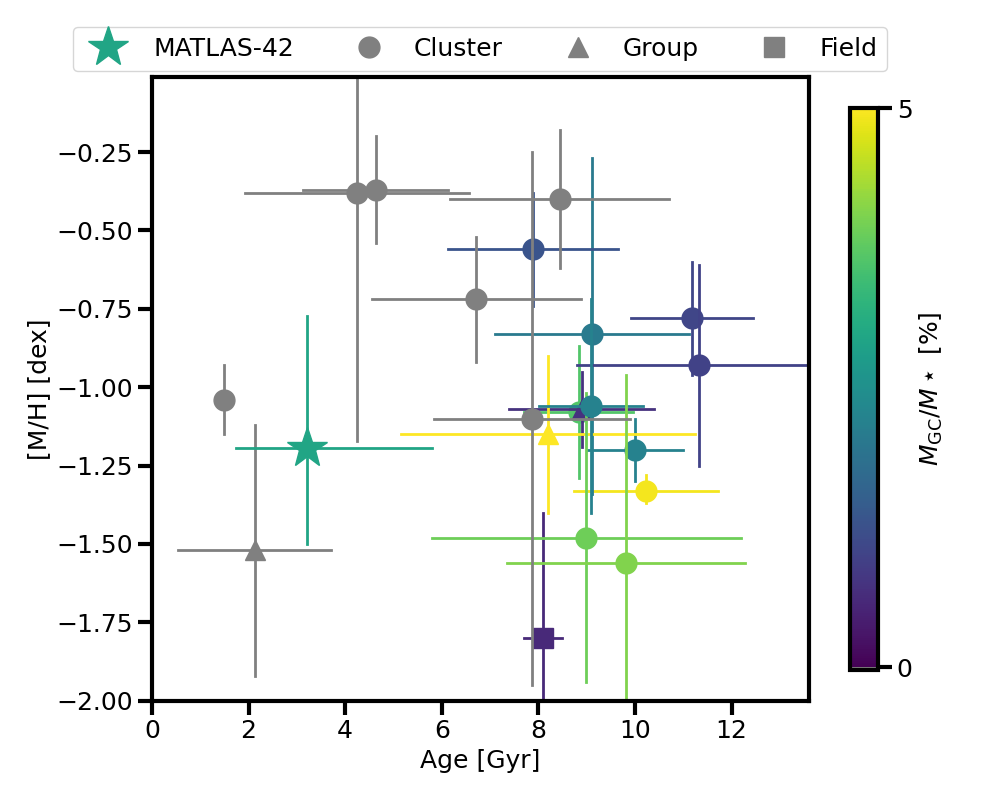}
    \caption{Mass-weighted metallicity vs mass-weighted mean stellar age. Point and colour styles follow Figure \ref{fig:mass_met}. In comparison to other UDGs known to be GC-rich, MATLAS-42 is noticeably younger. The young age may be attributable to it being in a less-dense group environment, while the other UDGs studied tend to reside in galaxy clusters.}
    \label{fig:age_met}
\end{figure}

In Figure \ref{fig:age_met} we place MATLAS-42 with the same UDGs in age--metallicity space. While there are a handful of UDGs as young as MATLAS-42, it is overall younger than many of the UDGs that have been studied spectroscopically thus far, including all with measured GC systems. It is worth noting MATLAS-42 is different from other GC-rich UDGs on two fronts. Firstly, it is young and contains gas, as noted above. Secondly, UDGs in lower-density environments are usually found to be GC-poor \citep{Somalwar2020, Jones2023}.

Given MATLAS-42 still contains significant gas, it is perhaps not unexpected that it hosts a young stellar population. It still leaves a puzzle as to why the other GC-rich UDGs are old and quenched while MATLAS-42 is not. It may be tempting to attribute the early quenching of most of the other UDGs to their being found in a cluster environment. However, it is worth noting that their phase space positioning within clusters does not generally align with the expectation that they were environmentally quenched \citep{Gannon2022, Forbes2023}. A separate quenching mechanism for at least some of these older UDGs is therefore required, with some attributing it to quenching from feedback as the GC system forms (e.g., \citealp{TrujilloGomez2022}). For MATLAS-42 the implication would be that this quenching mechanism was not as permanent as for the other GC-rich UDGs previously studied.

A slightly convoluted explanation for this difference may be that all GC-rich UDGs thus far studied quenched upon the initial GC formation, and the gas ejected from those UDGs now found in dense environments was lost to the galaxy cluster before it could resettle to the centre of its host dark matter halo. In this manner, the quenching time of a GC-rich UDG now found in a galaxy cluster may not exactly match its infall time onto the cluster as its gas was ejected by the GC system's formation and then was stripped by cluster infall before it was re-accreted. For MATLAS-42, the lack of environmental quenching would allow its gas to settle back into the centre of its halo following its initial ejection at the epoch of GC formation. This gas can form additional stellar mass, furthering the galaxy's enrichment to its current $[M/H]$.


To highlight the unusual properties of MATLAS-42 we can compare it to the four other group and field galaxies in the catalogue of \citet{Gannon2024b}, NGC~1052-DF2, NGC5846\_UDG1 (a.k.a., MATLAS-2019), UDG1137+16 and DGSAT~I. In comparison to both NGC5846\_UDG1 and DGSAT~I, which have a high fraction of their stellar mass in their GC system, MATLAS-42 is $\sim5$~Gyr younger than either. We conclude it is unlikely MATLAS-42 has experienced a similar star formation history to either galaxy. Despite this, \citet{Janssens2022} did note a knot of recent star formation within DGSAT~I, suggesting that it may be reigniting star formation and may soon be more similar to MATLAS-42 than it is currently, but as a whole, the galaxy is quiescent \citep{MartinNavarro2019}.

NGC~1052-DF2 is known to exhibit a population of abnormally bright GCs \citep{Shen2021b} and has been theorised to have formed from a bullet-like collision of dwarf galaxies $\sim8$~Gyr ago \citep{Silk2019, vanDokkum2022}. It is now embedded within a linear trail of galaxies projected on the sky \citep{vanDokkum2022, Keim2025, Tang2025}. MATLAS-42 does not exhibit a similar population of bright, monochromatic GCs \citep{Marleau2024}, and while there exist other nearby dwarf companions, MATLAS-41 and MATLAS-43, it is not clear that they follow a linear trail. That is, MATLAS-42 is not obviously embedded within a linear trail of galaxies on the sky. Given its younger age than NGC~1052-DF2, these galaxies should be much more tightly associated with one another than is seen for the galaxies in the NGC~1052 group. We conclude it is unlikely that MATLAS-42 is another example of this proposed exotic formation pathway. 

Finally, when compared to UDG1137+16, MATLAS-42 has a similar age and metallicity. Two differences remain: 1) UDG1137+16 is currently undergoing tidal stripping, although this may not be the cause of its large size \citep{Gannon2021}; 2) UDG1137+16 does not currently have its GC system measured. Given the other dwarf like properties of UDG1137+16 (i.e., its stellar mass, age and metallicity) it seems likely that this galaxy represents a ``puffy dwarf'' UDG, the tail of the normal dwarf galaxy size -- mass distribution to the largest sizes. As MATLAS-42 has a rich GC system, beyond that typically hosted by a dwarf galaxy, it would not easily fit with the picture of a ``puffy dwarf''. Based on our comparison to other currently studied field and group UDGs it is not obvious how this galaxy may have formed.

\subsection{MATLAS-42 Formation} \label{sec:formation}
The established formation pathway for many GC-rich UDGs is that of a ``failed galaxy'' or ``pure stellar halo'' \citep{vanDokkum2015, Peng2016, Danieli2022}. MATLAS-42 does not fit this formation pathway. Namely, it has not completely failed; its young age suggests that it has recently experienced star formation. Furthermore, it is not necessarily as GC-rich as many of the UDGs that best fit this pathway. For example, NGC~5846\_UDG1 has nearly 13\% of its luminosity within its GC system \citep[][although see \citealp{Muller2021} where a number closer to 5\% would be derived]{Danieli2022}. It is for this reason that \citet{Danieli2022} suggested that, once you account for the destruction of GCs with time, the galaxy would have had to form nearly its entire stellar mass within its GC system at formation. With $<3\%$ of its stellar mass within its GC system, it is very unlikely that the stellar body of MATLAS-42 is entirely/substantially comprised of disrupted GCs. Put another way, while a ``failed galaxy''-like formation pathway may have contributed to its initial evolution, it has now diverged from this pathway with subsequent star formation to be something else. This may hint at some level of diversity in evolution following the initial epoch of formation of ``failed galaxy'' UDGs.

Beyond the failed galaxy formation pathway, it has also been suggested that classical dwarf galaxies may become more GC-rich for their stellar mass in denser environments. This idea is commonly referred to as ``biasing'' (see e.g., \citealp{West1993, Peng2008}). MATLAS-42 is associated NGC~502, which is itself associated with NGC~524 \citep{Brough2006}. The total environmental halo mass is approximately $10^{13}$~M$_\odot$, based on the total GC numbers of NGC~524 itself \citep{Harris2013, Burkert2020}. MATLAS-42 is thus not in as dense an environment as those UDGs found in clusters, and the expected environmental enhancement to GC formation is not expected to be large. 

Thus far, we have largely assumed that both the `bluer' and `redder' populations of star clusters around MATLAS-42 are GCs. It is possible that one of the GC subpopulations is not old bona fide GCs (and hence $N_{\rm GC}$ is overestimated) but rather that it is comprised of young clusters with an age similar to the galaxy itself.  If these young clusters share the galaxy's metallicity, then they are likely to have blue colours. The old bona fide GCs would thus consist of only the red subpopulation, and $N_{\rm GC}$ would be reduced to around a dozen. The implication of the red subpopulation being the bona fide GCs is that there would have to be a large population of metal-rich GCs - more than what is traditionally seen for dwarf galaxies, where such metal-rich clusters are outliers in the underlying population (see e.g., the dwarf galaxy GC colour distributions in \citealp{Fielder2024}, fig. 7). Should these metal-rich clusters be confirmed for MATLAS-42, and/or further UDGs be found with significant populations of metal-rich GCs, it would be worthwhile to add this puzzle to the menagerie of oddities thus far discovered in UDG GC systems.

Even with a revision of the GC system downward to exclude a possible young star cluster population, MATLAS-42 would still be relatively GC-rich for its stellar mass. For example, \citet{Marleau2025} found that only $\sim$ 1/4 of the 1100 dwarf galaxy candidates in their EUCLID observations of the Perseus cluster had $>$2 GC candidates. Furthermore, we reiterate that of the 74 dwarf galaxies studied by \citet{Marleau2024} using the same methods, MATLAS-42 was clearly one of the most GC-rich, with 64\% of their galaxies shown to have few, if any, GCs. In having a dozen `bona-fide' GCs it would still be GC-rich within their sample and any formation scenario for MATLAS-42 would need to explain both star cluster populations.

GAMA 526784 \citep{Buzzo2025c} is similar to MATLAS-42 in that it exhibits two populations of star clusters, it has a gas reservoir, and it has been shown to have both an underlying old, quiescent stellar body, along with signs of recent star formation. Indeed, both galaxies show evidence of two components in their stellar body \citep{Marleau2024, Buzzo2025c}. While MATLAS-42 exhibited the majority of its star formation a few Gyr ago, for GAMA 526784, the period of its significant star formation was much more recent, with many of the star clusters having ages $<$1 Gyr. \citet{Buzzo2025c} posited that an established galaxy may experience localised cluster formation caused by gas instabilities, which are themselves the result of gravitational interactions from nearby companions. For MATLAS-42, this scenario seems entirely plausible.

However, galaxies can (and do) form stars without external triggering. For example, there is also the case of the ``Disco Ball'' Galaxy \citep{Khim2025}. This UDG hosts a significant population of star clusters (34$\pm$11) and shows signs of recent star formation activity, despite being on the red sequence. Its nearest spectroscopically confirmed neighbour is 1.7 Mpc away, suggesting that this recent star formation must not be triggered externally \citep{Khim2025}. It is currently unclear what the major driving processes are in the formation of the ``Disco Ball''. 

To best understand the reignition of star formation in MATLAS-42:

\begin{enumerate}
    \item Resolved HI and molecular gas observations would allow for an investigation into the distribution of its gas content, along with how it is turning its neutral gas into molecular gas for star formation.
    
    \item Deep spectroscopy targeting the ages and metallicities of individual star clusters around MATLAS-42 would help us better understand its star formation history. This would also aid in understanding the nature of its underlying `bluer' star cluster population.
\end{enumerate}

\subsection{Time Evolution of the GC system}

\begin{figure}
    \centering
    \includegraphics[width=0.48\textwidth]{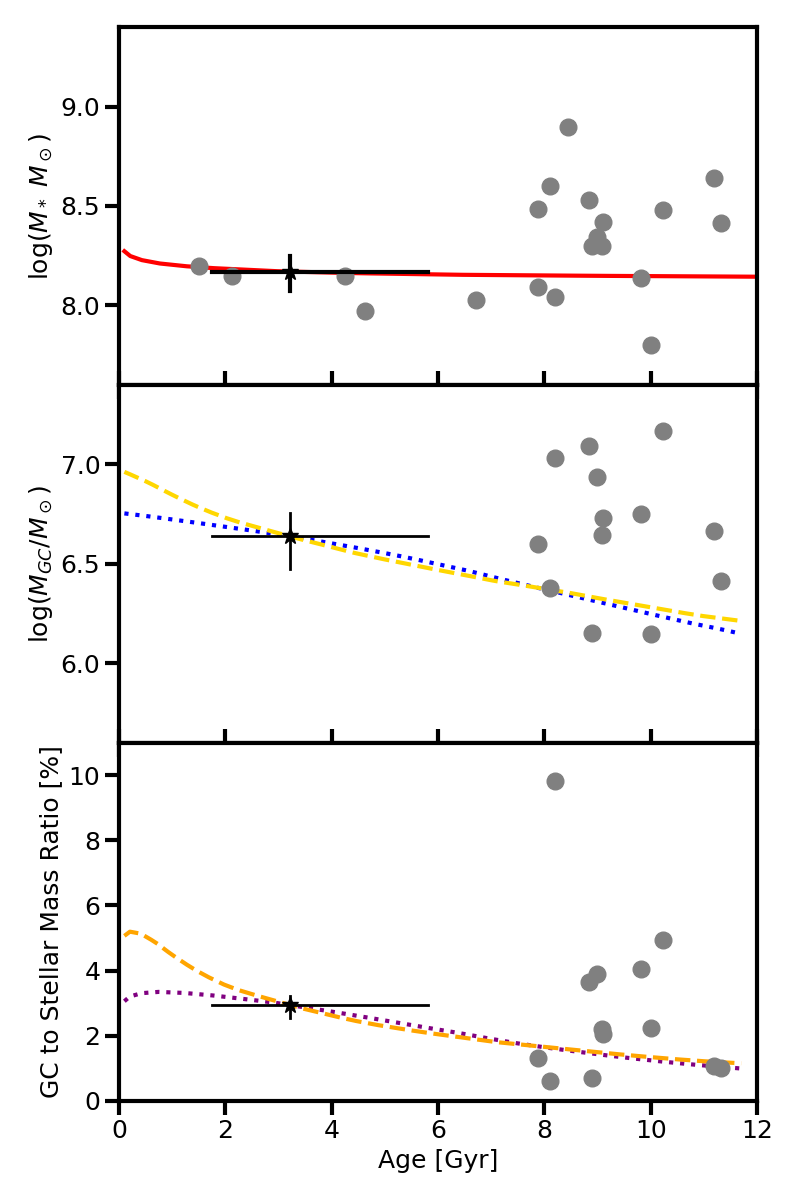}
\caption{A toy model for the possible evolution of the GC to stellar mass ratio of MATLAS-42. In all panels, we plot MATLAS-42 (black star) along with UDGs from the catalogue of \citet{Gannon2024b} with GC information (grey points). \textit{Top panel}: Stellar mass \textit{vs.} age. We include the model for the mass loss of a passively evolving \citet{Kroupa} IMF-based stellar population from the review of \citet{Courteau2014} as a red line. \textit{Middle panel}: Total GC system mass \textit{vs.} age. We include two models of GC system mass loss in time based on the simulations of \citet{MorenoHilario2024}. In blue (dotted line), we plot the median mass loss evolution coming from their galaxy `A' simulation, while in yellow (dashed line), we plot the median mass loss evolution coming from their galaxy `D' simulation. \textit{Lower panel}: GC to stellar mass ratio \textit{vs.} age. As this panel represents a ratio of the two panels above it, we combine the evolutionary curves from above for this panel. That is, the orange dashed curve is a combination of the red and yellow curves above, while the purple dotted curve is a combination of the red and blue above. When combining, the stellar mass lost in the GC system is gained by the stellar body of the galaxy. MATLAS-42 is plotted using our spectroscopic age, the stellar mass derived by \citet{Buzzo2024}, and a total GC system mass calculated using the GC counts of \citet{Marleau2024}, assuming an average GC mass of $2\times10^{5} M_\odot$. We conclude that while MATLAS-42 exhibits a rich GC system at the present day, this will decay with time. If it passively evolves, in $\sim7$~Gyr time (i.e., when it will be $\sim10$~Gyrs old similar to old UDGs), it will only have a GC system representing $\sim1$\% of its stellar mass. It is thus likely not a low redshift analogue of a ``failed galaxy' progenitor. }
    \label{fig:gc_evolution}
\end{figure}

Despite it being clear that MATLAS-42 has not formed and evolved via a similar pathway to the older, quenched, GC-rich UDGs, it might be tempting to perceive it as an analogue to how these UDGs may have looked at earlier times. To assess this, in Figure \ref{fig:gc_evolution} we build a toy model for the passive evolution of the GC-to-stellar mass ratio $M_{\rm GC}/M_\star$~with time (a similar quantity of GC specific mass, $S_{\rm M}$, is used in the literature; \citealp{Miller2007, Peng2008}). In the top panel, we show the passive evolution of stellar mass with time from the \citet{Kroupa} IMF models shown in the galaxy mass review of \citet{Courteau2014}. As galaxies passively evolve, stellar mass is slowly lost to stellar evolution. In the middle panel, we show the passive evolution of GC system mass from the simulations of \citet{MorenoHilario2024}. As GC systems passively evolve, individual GCs slowly evaporate and are disrupted, decreasing the total mass within the GC system. In this panel, we show two evolutionary curves from their galaxies `A' and `D', which encapsulate the variation of GC system mass evolution in their simulations. These galaxies also bracket the extremities of dynamical masses known for UDGs ($\sim 10^{9} - 10^{11} M_\odot$), with their galaxy `E' having a greater dynamical mass than has been measured in UDGs ($>10^{11} M_\odot$). In the bottom panel, we show the evolution of $M_{\rm GC}/M_\star$~with time. We create this toy model by combining the evolutionary curves of the above two panels, accounting for the fact that mass lost by the GC system is gained by the stellar body of the galaxy. Based on our toy model, we have a general expectation that the $M_{\rm GC}/M_\star$ ratio will slowly decrease with time for a passively evolving system. 

In terms of MATLAS-42, the implication is that its current ratio of $M_{\rm GC}/M_\star = 2.9\%$ for a galaxy that is 3.2 Gyr old will decrease to below 1\% within 7 Gyr. Put another way, if MATLAS-42 were to evolve passively until its mass-weighted age was 10 Gyr, as is the case for many of the quenched, GC-rich UDGs studied spectroscopically, it would have a noticeably lower $M_{\rm GC}/M_\star$ than they. Indeed, there are multiple UDGs in the \citet{Gannon2024b} catalogue that have $M_{\rm GC}/M_\star>5\%$. Given that our toy model demonstrates this ratio decreases with passive evolution, it is not possible to easily evolve MATLAS-42 into these systems with time. We conclude that MATLAS-42 should not be interpreted as a low redshift analogue of GC-rich UDGs' progenitors.

An assumption of our toy model is that there is no subsequent star formation within the galaxy. As MATLAS-42 currently contains HI-gas it is likely that this assumption is poor. Any subsequent star formation will form new stars; however, the star formation would also be expected to form GCs very inefficiently compared to star formation events at high-redshift, where the vast majority of GCs in the Universe have formed. Our toy model thus represents an upper limit of the evolution of $M_{\rm GC}/M_\star$~with time, with most future star formation generally expected to cause $M_{\rm GC}/M_\star$ to drop. 

A second assumption of our toy model is that the GC system of the galaxy is as young as the galaxy itself (i.e., 3.2 Gyr old), making it likely that they are younger star clusters than what is traditionally called a GC. Given that much of the mass loss of a GC system occurs when it is young, the star clusters that have ages more typical for a GC ($\sim$12 Gyrs) will experience less mass loss than our model suggests, possibly causing less decay in $M_{\rm GC}/M_\star$ than is modelled. However, as we have measured an age older than $\sim2$ Gyr we are beyond the point where much of the mass loss occurs for young star clusters. After this point the mass loss rate is relatively constant for either model. We suggest that this assumption has little impact on our results.

We reiterate the general conclusions of this subsection: 1) That our modelling suggests galaxies' $M_{\rm GC}/M_\star$ is generally expected to decline with time and 2) that MATLAS-42 is not a low redshift analogue of what a ``failed galaxy'' progenitor looked like at high redshift. It is not sufficiently GC-rich for this to be the case. At most, it may have formed similar to a ``failed galaxy'' but has recently rejuvenated star formation.

\section{Conclusions} \label{sec:conclusions}
In this work, we have studied the recessional velocity, stellar population and GC-population of the GC-rich, gas-rich, group UDG, MATLAS-42 using spectroscopic data from the Keck Telescopes. Our conclusions are as follows:

\begin{itemize}
    \item We measure a recessional velocity of 2433$\pm$8 km s$^{-1}$~ for MATLAS-42, confirming its association with the NGC~502 group and with the HI-gas previously assumed to have been associated with it. Confirmation of its environment/distance confirms its status as a UDG.

    \item We measure a mass-weighted age of $3.2^{+2.6}_{-1.5}$Gyr and a mass-weighted metallicity of [$M/H$]=$-1.19^{+0.42}_{-0.30}$ dex. This places MATLAS-42 at the lower, metal-poor edge of the regular mass-metallicity relationship for dwarf galaxies. It also places MATLAS-42 as being significantly younger than any other GC-rich UDG previously studied with spectroscopy.

    \item When considering established formation pathways and previously observed UDGs in low-density environments, we do not find an obvious formation pathway for MATLAS-42 and its rich GC system. Namely, it cannot be a simple ``failed galaxy'' as it has recently experienced star formation and is not quenched - a requirement of the formation pathway. If it formed via this pathway, it has now diverged from it. Further, its rich GC system cannot be explained by `biasing' as it does not reside in a particularly dense environment.

    \item Finally, we build a toy model of the evolution of $M_{\rm GC}/M_\star$. We find that its current $M_{\rm GC}/M_\star = 2.9\%$ will evolve to $<1\%$~in 7 Gyr of passive evolution. As such, MATLAS-42 cannot be a low redshift analogue of the progenitors of the previously studied quenched, GC-rich UDGs.
\end{itemize}

To best understand the galaxy further, measurements of its star formation history and gas content (to see when it rejuvenated) are required. Deep spectroscopy of its star cluster system would also help to disentangle the nature of its blue and red populations of star clusters. 

\section*{Acknowledgments}
We thank the anonymous referee for their work to improve the quality and clarity of this paper. DAF and JPB thank the ARC for financial support via DP220101863. AJR was supported by National Science Foundation grant AST-2308390. A.F.M. has received support from RYC2021-031099-I and PID2021-123313NA-I00 of MICIN/AEI/10.13039/501100011033/FEDER,UE, NextGenerationEU/PRT. Some of the data presented herein were obtained at Keck Observatory, which is a private 501(c)3 non-profit organization operated as a scientific partnership among the California Institute of Technology, the University of California, and the National Aeronautics and Space Administration. The Observatory was made possible by the generous financial support of the W. M. Keck Foundation. The authors wish to recognize and acknowledge the very significant cultural role and reverence that the summit of Maunakea has always had within the Native Hawaiian community. We are most fortunate to have the opportunity to conduct observations from this mountain.

\section*{Data Availability}
This data makes use of data from the Keck Telescopes which is available on the Keck Observatory Archive 18 months after the date of its observations. Per the request of \citet{Gannon2024b}, we thank the following authors for their contribution to the catalogue \citet{mcconnachie2012, vanDokkum2015, Beasley2016, Martin2016, Yagi2016, MartinezDelgado2016, vanDokkum2016, vanDokkum2017, Karachentsev2017, vanDokkum2018, Toloba2018, Gu2018, Lim2018, RuizLara2018, Alabi2018, FerreMateu2018, Forbes2018, MartinNavarro2019, Chilingarian2019, Fensch2019, Danieli2019, vanDokkum2019b, torrealba2019, Iodice2020, Collins2020, Muller2020, Gannon2020, Lim2020, Muller2021, Forbes2021, Shen2021, Ji2021, Huang2021, Gannon2021, Gannon2022, Mihos2022, Danieli2022, Villaume2022, Webb2022, Saifollahi2022, Janssens2022, Gannon2023, FerreMateu2023, Toloba2023, Iodice2023, Shen2023, Janssens2024, Gannon2024, Buttitta2025}. 



\bibliographystyle{mnras}
\bibliography{bibliography} 




\appendix

\section{\texttt{pPXF} Consistency Checks} \label{app:fitting}
\begin{figure}
    \centering
    \includegraphics[width=0.5\textwidth]{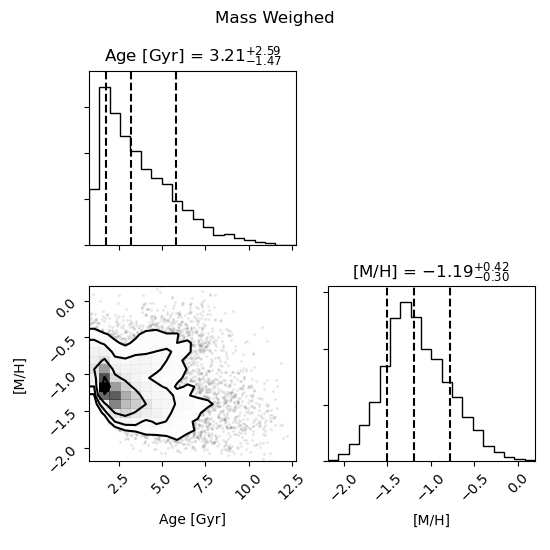}
    \caption{The mass-weighted results of our 10000 bootstrap fits of the spectrum. Ages and metallicities are relatively well-constrained by the data.}
    \label{fig:fits}
\end{figure}

As discussed in the main body of the text, while our bootstrapping of the spectra was able to derive a relatively stable age and metallicity for MATLAS-42 we were unable to derive a stable star formation history. We show our results for these fits in Figures \ref{fig:fits}. We find that the mass-weighted ages and metallicities are well-constrained. In order to test if our choice of fitting parameters greatly affected our results, we performed multiple different realisations of the spectra fitting. Namely, we refitted the spectra using combinations of with/without emission-line gas templates, 5th/10th/15th order multiplicative polynomials, adding in \texttt{pPXF}'s regularisation parameter as 0.01/0.05/0.1/0.5, and 7 different options of which regions of the spectra we masked. In all cases, ages were within our uncertainties. In nearly all cases, metallicities were within joint 1-$\sigma$ uncertainties with a slight bias to many realisations measuring MATLAS-42 to be slightly more metal-rich. Overall, the conclusions we draw in our paper would not be significantly affected were we to have chosen any of these consistency check bootstrap fits instead of our fiducial results. Finally, none of these tests resulted in well-constrained star formation histories. We conclude that, due to the young stellar population dominating much of the light in our spectrum, it is difficult to accurately assess the exact properties of the underlying older stellar population. As such, we do not include any discussion of the star formation history of MATLAS-42 in this work beyond taking note of it experiencing recent, significant star formation, and solely focus on its age and metallicity.

\bsp	
\label{lastpage}
\end{document}